\pdfoutput=1 % to force arxiv to use pdflatex
\documentclass[aps,prl,twocolumn,groupedaddress,amsmath]{revtex4}
\usepackage{comment}
\usepackage{graphicx}
\usepackage{amssymb}
\usepackage{bm}
\usepackage{amsmath,amsfonts,latexsym}

\newcommand*{\cA}{{\cal A}}

\newcommand*{\cZ}{{\cal Z}}

\newcommand*{\cT}{{\cal T}}

\newcommand*{\sgn}{\mathrm{sgn}}

\begin{document}

\title{Order parameter statistics in the critical quantum Ising chain}

\author{Austen Lamacraft} 
\author{Paul Fendley}
\affiliation{Department of Physics, University of Virginia,
Charlottesville, VA 22904-4714 USA}
\affiliation{All Souls College and the Rudolf Peierls Centre for Theoretical
Physics, University of Oxford, 1 Keble Road,  OX1 3NP, UK}
%\author{Paul Fendley} 
%\affiliation{Department of Physics, University of Virginia,
%Charlottesville, VA 22904-4714 USA;\\
%All Souls College and the Rudolf Peierls Centre for Theoretical
%Physics,\\ University of Oxford, 1 Keble Road,  OX1 3NP, UK}
\date{\today}
%\email{austen@virginia.edu}

\begin{abstract}

In quantum spin systems obeying hyperscaling, the probability distribution of the total magnetization
takes on a universal scaling form at criticality. We obtain this scaling function exactly for
the ground state and first excited state of the critical quantum Ising
spin chain. This is achieved through a remarkable relation to the partition
function of the anisotropic Kondo problem, which can be computed
by exploiting the integrability of the system.

\end{abstract}

\maketitle

The concept of \emph{universality} is central to our understanding of
continuous phase transitions. Universal physical quantities at or near
a transition in one system coincide with those in other systems that
share a few key characteristics, including dimensionality and symmetry
of the order parameter.  Continuous transitions are therefore grouped
naturally into \emph{universality classes} of common critical
behavior~\cite{domb1996cph}.  Familiar universal quantities include
critical exponents that characterize the singular behavior of
thermodynamic and response functions in the vicinity of the critical
point. Although the amplitudes of these singularities are entirely
system-dependent, certain \emph{amplitude ratios} are
universal~\cite{privman1991ucp}.

A particularly natural family of universal amplitude ratios
% that is particularly natural from the point of view of numerical
% simulation
is formed from the volume-integrated order parameter $M$ in a finite-size system by
%(e.g. the total magnetization of an Ising system)
%
\begin{equation}\label{A_def}
\cA_{2n}\equiv \frac{\langle M^{2n} \rangle}{\langle M^2 \rangle^n}\ .
\end{equation}
It is readily seen that the hypothesis of universality applied to this entire family is equivalent to the existence of a \emph{universal scaling function} $f(X)$ defined by
\begin{equation}\label{f_def}
f(X)\equiv s P(s X)
\end{equation}
where $s^2\equiv \langle M^2 \rangle$ is the variance of the order
parameter and $P(m)$ its probability distribution function. The
existence of this scaling function relating the probability
distribution for different system sizes is a consequence of
hyperscaling, as emphasized in Ref.~\onlinecite{Aji2001}.  Extensive
numerical work %in a variety of models
has confirmed the universality of  $s P(sX)$
\cite{bruce1981pdf,binder1981fss}.  For the important benchmark of
the two-dimensional classical Ising model, Ref.~\onlinecite{Salas2000}
is a useful guide to the literature as well as a tour-de-force
numerical study.

The aim of this work is to remedy two striking deficiencies in our
present understanding of order parameter distributions: the small number of
analytical results and the scant attention that has been paid to
\emph{quantum} phase transitions. The study of order parameter
fluctuations at quantum critical points, aside from being inherently
interesting, is motivated by recent experiments in atomic physics,
where the measurement of the full distribution of global fluctuating
observables has become a
reality~\cite{gritsev2006fqd,hofferberth-2007}.
For the quantum Ising chain, we compute $P(m)$ exactly by
relating its generating function to the partition function of a
particular anisotropic Kondo problem. This remarkable relationship
allows the application of the powerful analytic methods developed to
solve this and other quantum impurity problems.

The quantum Ising chain, often referred to as the
transverse-field Ising model, has Hamiltonian 
%\cite{kogut1979}
%TFIM Hamiltonian~\cite{sachdev_book}
%
\begin{equation}\label{TFIM}
H=-\sum_{i=1}^L \left[h\sigma_i^x+\sigma_i^z\sigma_{i+1}^z \right]\ .
\end{equation}
For the moment, we impose periodic boundary conditions so that
$\sigma^z_{L+1}\equiv\sigma^z_{1}$. 
%As we will see explicitly,
%the distribution function is sensitive to the boundary
%conditions. 
The critical point of the model Eq.~(\ref{TFIM}) is at
$h=1$, separating the ordered ($h<1$) and disordered ($h>1$)
phases. The magnetization
$M\equiv \sum_i \sigma_i^z/2$
does not commute with $H$, so eigenstates of $H$ are typically sums
over states with different eigenvalues of $M$. The distribution
functions of the magnetization in the ground states for various
boundary conditions 
were studied numerically in Ref.~\onlinecite{Eisler2003}, and
good scaling of $s P(sX)$ onto a universal curve was found for
$L\gtrsim 16$. There are also analytical results for the distribution
of the \emph{transverse} magnetization $ \sum_i
\sigma_i^x$~\cite{Cherng2007,Eisler2003}.
%Namely, the magnetization
%distribution of the quantum Ising chain is the Fourier transform of
%the partition function of a quantum impurity problem.

The magnetization distribution in
a state $|\alpha\rangle$ is 
\begin{equation*}
P_\alpha(m)=\langle \alpha|\delta(m-M) |\alpha \rangle  .
\label{Pdef}
\end{equation*}
Rewriting the delta
function as an integral
%using $\delta(x)=\int_{-\infty}^\infty d\lambda e^{i\lambda x}$
gives
\begin{equation*}
P_\alpha(m)=\int_{-\infty}^\infty \frac{d\lambda}{2\pi} e^{-i\lambda m}
\chi_\alpha(\lambda)
%\langle \alpha| e^{i \lambda M} |\alpha \rangle \ .
\end{equation*}
%
%Thus $P_\alpha(m)$ is the Fourier transform of the generating function
where $\chi_\alpha(\lambda) \equiv \langle \alpha| e^{i \lambda M}
|\alpha \rangle $ is the generating function of the moments of the
distribution.
%of the moments of the distribution
%$$\langle \alpha |M^{2n}|s\rangle = \int_{-\infty}^\infty dm\, P_\alpha(m) m^{2n}
%L\ .$$
%The arguments of \cite{Aji2001} (adapted to the quantum case)
%imply that the distribution function $P_0(m)$
%for the ground state becomes a continuous scaling
%function in the $N\to\infty$ limit {\tt do they -- to think about}. 
%By explicit computation, we
%will confirm this, and show that $P_\alpha(m)$ for several excited states
%are also scaling functions.
%A few properties of $P_\alpha(m)$ follow on general grounds.  
%\begin{itemize}
%\item 
The flip
operator ${\cal F}\equiv \prod_i \sigma_i^x$ commutes with the
Hamiltonian, so the resulting ${\mathbb Z}_2$ symmetry requires that
$P_\alpha(m)=P_\alpha(-m)$. 
%\item 
A generalized
Lee-Yang theorem shows that the generating function for the ground
state has the factorization \cite{newman1975}
\begin{equation}\label{chi_factor}
\chi_0(\lambda)=\prod_p\left(1-\frac{\lambda^2}{E_p}\right)
\end{equation}
 for real positive $\{E_p\}$. The even cumulants, given by the coefficients of the expansion of $\ln \chi_0(\lambda)$, are then
$$\langle M^{2n}\rangle_c=(-1)^{n-1}\frac{(2n)!}{n}\sum_p (E_p)^{-n}
$$
and therefore alternate in sign. 
%A by-product of our work will be an explicit identification of the $\{E_p\}$.

%\end{itemize}

In the scaling limit, the sum in the operator $M$ can be replaced with
an integral $M=\int_0^L dx\, \sigma(x)$, where $\sigma(x)$ is the
standard Ising quantum field. Expectation
values in the ground state can be computed in the path-integral
picture by taking Euclidean spacetime to be a very long cylinder of
circumference $L$; the long cylinder means that in the Hamiltonian picture the system is projected onto its ground
state. The path integral for the generating
function for the ground state is then
%$$\chi_0(\lambda) = \int[D\sigma(x,t)] e^{-S[\sigma]} e^{i\lambda \int_0^L
%  dx \sigma(x,0)} $$ where $S[\sigma]$ is the action of the Ising
%field theory over the spacetime cylinder.  This is
precisely the partition function of the 2d classical Ising model with
an imaginary magnetic field along a defect line wrapping around the
cylinder. In this path integral, we are free to exchange the roles of
space and time so that the new Euclidean ``time'' direction $\tau$ is
periodic. Since all the operators in $\chi_0$ were originally at the
same time, this exchange puts them all at the same spatial
position. Thus in this new picture, $\chi_0$ describes the continuum
limit of infinitely-long Ising chain at temperature $1/L$ and an
imaginary magnetic field $i\lambda$ at point $0$. The underlying
lattice Hamiltonian is that of the quantum Ising
chain (\ref{TFIM}) on an infinite line, so that
the sum runs from $i=-\infty$ to $\infty$.  The generating function is
simply
\begin{eqnarray}\label{gen_fn}
\chi^{}_0(\lambda)=\hbox{tr } \left[ e^{-H L} \cT\, e^{i\lambda\int_0^{L} d\tau\,
\sigma^z_0(\tau)} \right]
\end{eqnarray}
where $\cT$ represents time-ordering and the (Euclidean) time
dependence of the operator denotes the Heisenberg picture:
$\sigma_0^z(\tau)=e^{H \tau}\sigma_0^z e^{-H\tau}$.

%The resulting integrals at low orders in the
%classical two-dimensional case have been studied intensively
%numerically \cite{Salas2000}.  Here we will follow a different
%strategy, mapping $\chi_0$ to the partition function of a quantum
%impurity problem which can be computed exactly.

Writing $\chi^{}_0$ as (\ref{gen_fn}) allows us to compute it exactly
at the critical point.  We first relate $\chi^{}_0$
to the partition function of a famous quantum impurity model,
the anisotropic Kondo problem \cite{anderson1969}.  We then apply the
methods of integrability to compute $\chi_0(\lambda)$ and hence
$P_0(m)$.

There are two ways of showing why the Kondo problem arises. The first
is quite direct. At the critical point $h=1$, the trace in
Eq.~(\ref{gen_fn}) can be expressed as an expectation value in the
critical Ising field theory. We then can expand in $\lambda$ and use the
known spin correlation functions in the field theory to
write integral expressions for the moments. The integrals for the term
order $\lambda^{2n}$ term are over values of $0<\tau_j<L$, but because
spin correlations are independent of the ordering of the $\tau_j$, we
can multiply by $(2n)!$ and order them
$0<\tau_1<\cdots\tau_{2n}<L$. This allows us to exploit a marvelous
result for critical Ising correlators when all spin fields lie on a
cycle of a cylinder \cite{kadanoff1971doa}:
%of Eq.~(\ref{pert_exp}) is proportio
%
\begin{equation}\label{KC_corr}
\langle \sigma(0,\tau_{2n})\cdots\sigma(0,\tau_1)\rangle
\propto 
\prod_{i>j}^{2n}\left[\sin\left(\pi\frac{\tau_i-\tau_j}{L}\right)\right]^{(-)^{i+j}2g}
\end{equation}
%
%\begin{figure}
%\centering  \includegraphics[width=0.4\textwidth]{cylinder.pdf}
%\caption{Representation of the correlation function Eq.~(\ref{KC_corr}) as a string of antiferromagnetic defects
 %\label{fig:cylinder}}
%\end{figure}
%
where $g=1/8$, the scaling dimension of the spin field.
This formula is valid only when the $\tau_j$ are
ordered.  
Keeping track of the constants in front of (\ref{KC_corr})
shows $\chi_0(\lambda)$ is a function of the
dimensionless quantity $z\equiv\lambda (2\pi a)^{1/8}L^{7/8}$, which
establishes the system size independence of the amplitude ratios
Eq.~(\ref{A_def}) and the scaling form of the order parameter
distribution (Eq.~(\ref{f_def})) for this model, with $s \propto
a^{1/8}L^{7/8}$.

%$\cN_n\equiv \left(\pi a/L\right)^{2ng}$, with $a$ some microscopic cut-off. 
%On changing boundary conditions we have $\Delta\varphi_0=\pi/2$, so that  the exponent appearing in %Eq.~(\ref{KC_corr}) is $1/4$. $g=1/8$ is  
%With this realization, it is possible to use Jordan-Wigner fermions or
%the conformal field theory results of Ref.~\cite{oshikawa1997} to
%obtain the correlation function that appears as t

The correspondence with the Kondo problem is now apparent.
$\chi_0(\lambda)$ takes the form of the partition function at
imaginary fugacity $iz$ of a Coulomb gas on a ring.
The gas consists of positive and
negative charges with logarithmic interactions; 
because of the
$(-1)^{i+j}$ in (\ref{KC_corr})
the signs of the charges required to alternate in space. This is precisely the celebrated
Anderson-Yuval expansion for the partition function $\cZ_K$ of the
Kondo model, describing the interaction of a spin-1/2 impurity with a Fermi
gas~\cite{anderson1969}. The expansion in $z$ is a
perturbative expansion in the spin-flip part of the Hamiltonian, with
alternation in signs of charges 
arising from the two spin
states of the impurity. In
the Kondo problem $g$ parametrizes the anisotropy (in spin space)
of the interaction. Thus we have shown that
\begin{equation}
\chi_0(\lambda) = \frac{1}{2} \cZ_K(iz),
\label{chiZ}
\end{equation}
with $g=1/8$. A very similar result was derived for the chiral version
of this problem, describing a point contact in a $p+ip$ superconductor
\cite{fendley2006ddb}, and is in accord with the results for the
boundary entropy of the Ising model with a defect magnetic field
\cite{oshikawa1997,leclair1999}.

An illuminating way of rederiving Eq.~(\ref{chiZ}) is to
use boundary conformal field theory. % The corresponding
%field theories have their spin fields
%$\sigma^{(1)}(0,\tau)=\sigma^{(2)}(0,\tau)$ identified at the boundary as
%well. 
%\end{comment}
Expanding Eq.~(\ref{gen_fn}) in
$\lambda$ and using the time ordering gives
\begin{widetext}
\begin{eqnarray}\label{pert_exp}
\chi_0^{}(\lambda)=\sum_{n=0}^\infty(i\lambda)^{2n}
\int_0^L d\tau_n \int_0^{\tau_n} 
d\tau_{n-1}\cdots \int_0^{\tau_2} d\tau_1
\mathrm{tr} \Big[
 e^{-H(L-\tau_{2n})}\sigma_0^z e^{-H(\tau_{2n}-\tau_{2n-1})}\sigma_0^z\cdots
\sigma_0^z e^{-H(\tau_2-\tau_1)}\sigma_0^z e^{-H\tau_1}\Big]
\end{eqnarray}
\end{widetext}
Since $\sigma^z \sigma^x \sigma^z = - \sigma_x$, the effect of the
boundary magnetic field is to flip the sign of the transverse field
$h$ at site zero for {\em every other} interval between insertions of
$\sigma_0^z$.  This means we can effectively treat $h$ as being
time-dependent, i.e.\ $h(\tau)=h\prod_{j=1}^{2n}\sgn(\tau_{j}-\tau)$.
The Ising spin chain has only nearest-neighbor interactions, so we can
``fold'' the theory in half at site $0$, turning the defect into a
boundary.
The continuum analog of 
$\sigma^z_0(\tau)$ is then a {\em boundary-condition-changing
  operator} ${\cal B}(\tau)$ 
\cite{cardy1989}. Here, inserting 
${\cal B}(\tau_j)$ at an
instant $\tau_j$ toggles between two
different critical boundary conditions. 
The bosonization analysis of Ref.~\cite{oshikawa1997} allows us to show that
${\cal B}(\tau)$ is identical to the spin-flip operator in the
anisotropic Kondo model. This follows from two key facts about 
${\cal B}(\tau)$: it has dimension $1/8$, and inserting it simply toggles back and forth between two Dirichlet-type boundary conditions on the boson, corresponding to fixed values $\varphi=\pi/4$ and $\varphi=3\pi/4$ of the boson field. A free boson calculation immediately yields Eq.~(\ref{KC_corr}) with $2g=\Delta\varphi^2/\pi^2=1/4$.

\begin{figure}%[ht]
\begin{center}
\includegraphics[width=0.4\textwidth]{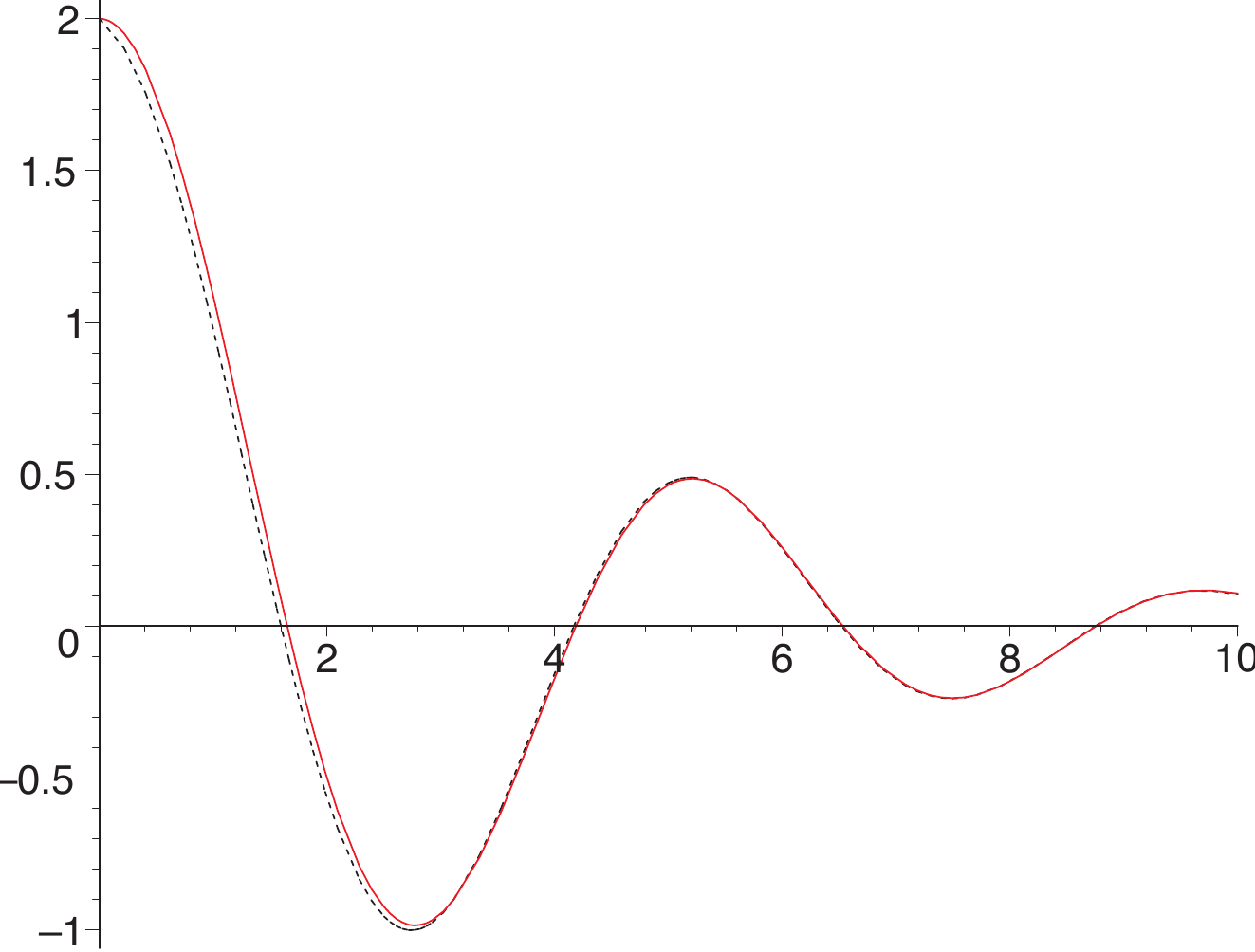}
{
    \put(-223,112){$\cZ_K(i\lambda)$}
    \put(0,30){$\lambda$}
}
\caption{$\cZ_K(i\lambda)$ at $g$=1/8 and
  (dashed) its approximation (\ref{Kasym})
 \label{fig:Zk}}
\end{center}
\end{figure}
We now find $\chi_0(\lambda)$, its asymptotics, and its
moments. There are three distinct ways of evaluating the partition function
$\cZ_K(iz)$, all of which work to high numerical
accuracy. One way is to use the thermodynamic Bethe ansatz
\cite{Tsvelik1983,Fendley1996}, the second is to use series expansions
\cite{Fendley1995,Fendley1996}, and the third is to compute the
spectral determinant of an associated ordinary differential equation
\cite{Dorey1999,Bazhanov1999isc}. We have used the latter two
approaches, both of course giving the same
result, displayed in Fig.~\ref{fig:Zk} along with a very accurate
asymptotic expression to be discussed below.
In accord with the generalized Lee-Yang theorem mentioned earlier, the
zeroes $\cZ_K(iz)$ occur at real $z$.
In fact, the set $\{E_p\}$
appearing in Eq.~(\ref{chi_factor}) are just the eigenvalues of the
spectral problem. 
%Since the ``physical'' Kondo partition function corresponds to
%$i\lambda$ real, there is no reason why it cannot be zero for
%$\lambda$ real, and indeed we see zeros along the axis.
%

\begin{figure}%[ht]
\centering  \includegraphics[width=0.5\textwidth]{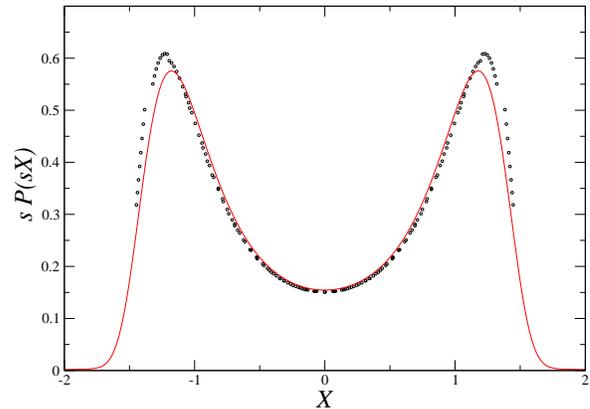}
\caption{Dots show numerical data for $sP(sX)$ for the ground state of the critical Ising chain for $L=16-23$. The red line is the scaling function $f_0(X)$ for the ground state, the Fourier transform of $\cZ_{K}(iz)$.
 \label{fig:histogram}}
\end{figure}
The scaled
distribution function $sP(sX)$ is
obtained by
(numerically) taking the Fourier transform of $\cZ_K(iz)$, and shown in
Fig.~\ref{fig:histogram}.
We have tested our
predictions by exact diagonalization of the lattice Hamiltonian
Eq.~(\ref{TFIM}) for $L=16-23$ using the ALPS
libraries~\cite{alet2005apo,Troyer1999}. 
We find excellent agreement, with the numerical results getting closer
to the exact curve as $L$ is increased.

We find accurate approximations for the scaling functions and
exact values of the moments
by utilizing the functional relation
between the partition
functions of the Kondo and the boundary sine-Gordon models
\cite{Bazhanov1996,fendley1995a,Fendley1996}
\begin{equation}\label{Z_relation}
\cZ_{K}(2i\sin(\pi g)z)=\frac{\cZ_{BSG}(e^{i\pi g} z)+\cZ_{BSG}(e^{-i\pi g} z)}{\cZ_{\mathrm{BSG}}(z)}
\end{equation}
Both partition functions
have similar Coulomb-gas expansions,
but $\cZ_{BSG}$ is given in terms of  \emph{unordered}
charges. The asymptotic
expression for $\cZ_{BSG}(z)$, valid for large $z$ in a region near
the positive real axis is \cite{Fendley1996} 
$$
\cZ_{BSG}(z) \approx \exp\left(\frac{\sqrt{\pi}\,\Gamma({\alpha/\pi})}
{\Gamma(\frac{1}{2}-\frac{\alpha}{\pi})}
\left(\frac{z}{\Gamma(g)}\right)^{1/(1-g)}\right)\ 
$$ where $\alpha= \pi g/(2-2g)$. (In the differential equation approach
\cite{Dorey1999,Bazhanov1999isc}, this expression arises from the WKB
approximation.)  Plugging this into (\ref{Z_relation}) gives
\begin{equation}
\cZ_{K}(i\lambda) \approx 2\cos\left[\cos(\alpha)\,(m_0\lambda)^{\frac{1}{1-g}}\right]
e^{-\sin(\alpha)\,(m_0\lambda)^{1/(1-g)} }
\label{Kasym}
\end{equation}
where 
$$m_0=\left(\frac{2\sqrt{\pi}\,\Gamma(\frac{1}{2} - \frac{\alpha}{\pi})}
{\Gamma(1-\frac{\alpha}{\pi})} \right)^{1-g} 
\frac{\Gamma(1-g)}{2\pi}\ .
$$ As $\lambda\to\infty$ along the real axis, this falls off as an
oscillating exponential. Even though this asymptotic expression in
principle holds only at large $\lambda$, when $g=1/8$ it is a quite
accurate approximation. It is plotted along with the exact expression
in Fig.~\ref{fig:Zk}.  Taking the Fourier transform of the asymptotic expression, one finds
$$P_0(m)\approx \frac{2\pi(1-g)}{m_0} 
\sum_{k=0}^{\infty} \frac{1}{(2k)!\,\Gamma((2k+1)g-2k)}
\left(\frac{m}{m_0}\right)^{2k}
$$
while a stationary phase approximation to the Fourier integral
gives $f_0(X\to\infty)\propto 
X^{-1+1/2g}\exp\left(-cX^{1/g}\right)$ with $c={g} [m_0/(1-g)]^{1-1/g}$. 

The moments of the distribution are related to $\cZ_K$ by
$\langle M^{2n}\rangle\propto (2n)!K_{2n}/2$, where
\begin{equation*}
\cZ_K(z) = 2 + \sum_{n=1}^\infty K_{2n}z^{2n}, 
\qquad
\cZ_{BSG}(z) = 1 + \sum_{n=1}^\infty B_{2n}z^{2n}\ .
%\label{kondopert}
\end{equation*}
The 
universal amplitude ratios are therefore
\begin{equation*}
\cA_{2n}=\frac{(2n)!}{2}\frac{K_{2n}}{(K_2)^n}
\end{equation*}
The $B_{2n}$ have explicit series expressions \cite{Fendley1995} which
allow accurate numerical evaluation; the $K_{2n}$ are then found 
in terms of the $B_{2m}$ with $m\le n$ 
using
Eq.~(\ref{Z_relation}). 
The first two ratios are exactly $\cA_{4}=1.43138\dots$ and
$\cA_{6}=2.35464\dots$, while from exact diagonalization
we find
\begin{table}[h]
\begin{tabular}{|c||c|c|c|c|c|c|c|c|}
\hline
$L$&16&17&18&19&20&21&22&23\\
\hline
$\cA_{4}$ &1.390& 1.393& 1.395& 1.396& 1.398 &1.399&  1.401& 1.402
\\
$\cA_6$ & 2.163&  2.173&  2.182&  2.190&  2.198&  2.204&  2.211&  2.216
\\
\hline
\end{tabular}
%\caption{Exact diagonalization data for the first two non-trivial
% moment ratios $\cA_{2n}$ as a function of system size.\label{tab:data}}
\end{table}

\begin{figure}%[ht]
\centering  \includegraphics[width=0.5\textwidth]{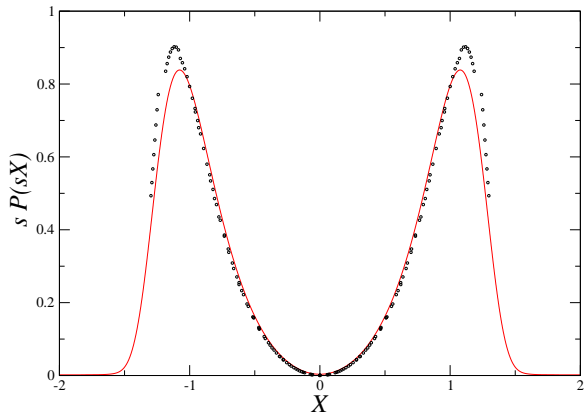}
\caption{Comparison of numerical and analytical results for the first excited state scaling function $f_1(X)$.
 \label{fig:first_ex}}
\end{figure}
This method can be extended to the computation of the distribution
function for excited states as well. The lowest-lying excited state
can be viewed as the ground state of a periodic
Ising system with an antiferromagnetic defect. In the Hamiltonian
formulation of the field theory, this excited state is given by acting
on the ground state with the spin field, so the generating function
for the moments of the first excited state is given by inserting spin
fields $\sigma(x=+\infty,\tau)$ and $\sigma(x=-\infty,\tau)$ into the
correlator (\ref{KC_corr}).
These insertions amount to setting $p=1/4$
%in additional factors $2\cos((4g\pi/L)\sum_i (-1)^i\tau_i)$ and the
%corresponding partition functions may be found by the same methods. 
in the conventions of \cite{Fendley1995} or $p=1/8$ 
in \cite{Bazhanov1996}.  The resulting
scaling function $f_1(X)$ of the first excited state is shown in
Fig.~\ref{fig:first_ex}. It is quite different from that of the ground
state, illustrating the importance of different boundary conditions.

We would like to thank Fabian Essler for useful discussions, Viktor
Eisler for sharing the numerical data from Ref.~\cite{Eisler2003}, and
John Cardy for pointing out the history of the relation
Eq.~(\ref{KC_corr}).  This research has been supported by the NSF
under grants DMR-0412956 and DMR/MSPA-0704666, and by an EPSRC grant
EP/F008880/1.

%\bibliographystyle{apsrev}
%\bibliography{OP_fluct} 

\end{document}